\documentclass{article}

\usepackage{fullpage}
\usepackage{bibunits}
\usepackage{graphicx}
\usepackage{amsmath}
\usepackage{amssymb}
\usepackage{hyperref}
\usepackage{cleveref}
\usepackage{siunitx}

\graphicspath{{../fig/}{fig/}{..}}
\newcommand{\figbf}[1]{\uppercase{\textbf{#1}}}
\newcommand{\key}[1]{\raisebox{0.5pt}{\protect\includegraphics{key/#1}}\,}

\title{\Large Optimal Navigation in Microfluidics via the Optimization of a Discrete Loss}

\author{
  Petr Karnakov\thanks{These authors contributed equally.} \and
  Lucas Amoudruz\footnotemark[1] \and
  Petros Koumoutsakos\thanks{Corresponding author: \href{mailto:petros@seas.harvard.edu}{petros@seas.harvard.edu}}
}

\date{\small%
  Computational Science and Engineering Laboratory, Harvard University, Cambridge, MA 02138, USA\\[1ex]
  January 29, 2025
}

\begin{document}

\maketitle

\renewcommand{\thefootnote}{}
\footnotetext{© 2025. American Physical Society. This is the author's version of the work. The official published article is available at \url{https://doi.org/10.1103/PhysRevLett.134.044001}.}
\renewcommand{\thefootnote}{\arabic{footnote}}

\begin{abstract}
Optimal path planning and control of microscopic devices navigating in fluid environments is essential for applications ranging from targeted drug delivery to environmental monitoring.
These tasks are challenging due to the complexity of microdevice-flow interactions.
We introduce a closed-loop control method that optimizes a discrete loss (ODIL) in terms of dynamics and path objectives.
In comparison with reinforcement learning, ODIL is more robust, up to three orders faster, and excels in high-dimensional action/state spaces, making it a powerful tool for navigating complex flow environments.
\end{abstract}

\section{Introduction}

The navigation and control of microfluidic devices are central in numerous fields, including precision medicine applications~\cite{barbolosi2016computational}, micro-manipulation~\cite{yun2013cell,ye2012micro,basualdo2022control}, targeted drug delivery~\cite{liebchen2019,amoudruz2022}, and  biomedical applications~\cite{bunea2020recent}. Nevertheless, there are numerous challenges in ensuring  precision and efficiency of transport in highly viscous flows ~\cite{safdar2017light}.
In this letter, we introduce a novel method for effective navigation and control in microfluidic environments that integrates the optimization of a discrete loss (ODIL)~\cite{karnakov2024odil} with neural networks (NNs) and automatic differentiation techniques.
ODIL relies on casting the discrete form of the governing equations as a minimization problem.
Here, ODIL builds upon the direct collocation method, a widely used technique in the optimal control community~\cite{kelly2017introduction,bordalba2022direct}, but uses NNs to represent the policy.
The use of NNs in the direct collocation method is not common due to the high computational cost associated with casting navigation as a constrained optimization problem.
Indeed, these methods rely on numerical linear algebra with matrix decomposition~\cite{betts2010practical}, which makes the method expensive as the number of parameters increases.

The use of NNs in optimal control has been explored since the 90s~\cite{chen1990back,hunt1992neural}.
Recent advances include applications  to nonlinear control~\cite{adhau2021constrained,jin2020pontryagin} and model predictive control~\cite{amos2018differentiable,karg2020efficient}, employing differentiable control policies and leveraging modern computational tools like automatic differentiation (AD)~\cite{baydin2018automatic,cao2005formulation}.
The adjoint approach for nonlinear optimal control~\cite{sandoval2022neural} has been suggested in the context of Neural ODEs~\cite{chen2018neural,liu2021second,katrutsa2024data}.
However, despite advances, there are limitations in differentiable control that hinder its effectiveness in solving flow navigation problems.
The target information enters the objective function but it is not tightly coupled with the system dynamics which can lead to sub-optimal solutions.
Moreover, the deployed explicit integrators may lead to instabilities, further challenging the controllers.
We also note that differentiable solvers and adjoint-based techniques assume that the forward problem is well-posed.
This may not always be the case as evidenced by the ``notorious test problem'' in optimal control~\cite{betts2010practical,bock1983recent},
where the forward solution is unstable, but adding the knowledge of the target position regularizes the problem.

Sampling-based algorithms~\cite{karaman2011sampling,arslan2017sampling} for optimal motion planning offer a powerful approach for complex robotic tasks, particularly in high-dimensional or complex environments. However, they are limited by slow convergence to optimal solutions, potential difficulties in narrow passages, sensitivity to parameters, and are thus not well suited for complex environment, and real-time or near-optimal solutions. To the best of our knowledge, these methods have not found any applications in fluid flows, except for recent work~\cite{liu2024adaptive} that combines adaptive sampling methods with physics informed NNs (PINNs).

Reinforcement learning (RL)~\cite{sutton2018reinforcement} is gaining attention as a potent approach to tackle navigation problems in mobile robots~\cite{zhu2021deep}, and for modeling and controlling complex fluid flows~\cite{reddy2018glider,verma2018efficient,gunnarson2021learning,colabrese2017,biferale2019,nasiri2022,amoudruz2022,amoudruz2024path,zhu2022point,li2019deep}.
RL uses neural parametrization of policies within Markov decision processes and has shown promising results, especially with the integration of deep learning techniques~\cite{mnih2013playing,verma2018efficient,novati2019}.
Furthermore RL produces robust policies due to its closed-loop formulation~\cite{thuruthel2018model,zavoli2021reinforcement,amoudruz2024path}.
More recently extensions of RL to high dimensional problems have shown great promise in solving complex tasks~\cite{komorowski2018artificial,kendall2019learning}.
However, RL may involve extensive tuning of hyper-parameters and heuristics in reward shaping, and often results in computationally costly implementations.

The proposed ODIL framework~\cite{karnakov2024odil} addresses several of the challenges mentioned above by leveraging automatic differentiation and standard gradient-based optimization methods~\cite{kingma2014adam}, available in many modern machine learning (ML) packages.
Moreover, it deploys a custom multigrid technique for accelerated convergence~\cite{karnakov2023flow} offering a fast and convenient tool to solve navigation and control problems for microfluidics.
We demonstrate the effectiveness of the ODIL approach on a series of benchmark navigation and control problems in microscale flow environments.
In comparison to RL, which currently represents the state-of-the-art for path planning and control problems in these applications~\cite{colabrese2017,biferale2019,nasiri2022,amoudruz2022,mo2023chemotaxis,mo2023challenges,amoudruz2024path}, ODIL is more robust, in particular for high-dimensional state/action spaces while requiring between one to three orders of magnitude fewer policy evaluations.

\section{Methods}
We consider path planning problems described by optimization of the form
\begin{gather}
  \label{eq:time}
  \text{minimize } T \\
  \label{eq:ode}
  \text{subject to }\dot{\mathbf{x}} = \mathbf{f}(\mathbf{x}, \mathbf{a}(\mathbf{x})),
  \quad t\in(0,T),
  \\
  \label{eq:constraints}
  \mathbf{x}(0) = \mathbf{x}_\text{start},
  \quad \mathbf{x}(T) = \mathbf{x}_\text{target},
\end{gather}
where $\mathbf{x}(t)$ represents the state of the system,
$\mathbf{a}$ is the control policy,
$\mathbf{f}$ describes the dynamics of the system,
$T$ is the travel time,
$\mathbf{x}_\text{start}$ is the initial state,
and $\mathbf{x}_\text{target}$ is the target state.
The task is to find a policy $\mathbf{a}\in A$ that minimizes the travel time
$T$ under the constraints (\cref{eq:ode,eq:constraints}),
where $A$ is a case-dependent feasible set of policies.
We represent the policy with a fully connected NN with weights $\theta$, $\mathbf{a} = \mathbf{a}_\theta$.
The NN consists of two hidden layers of size $128\times 128$, $\tanh$ activation in the hidden layers, and a case-specific activation function in the output layer.
For brevity, we assume that the right-hand side $\mathbf{f}$ and the policy $\mathbf{a}_\theta$ do not explicitly depend on time $t$.

\section{ODIL approach}
The ODIL approach reduces the constrained minimization problem (\cref{eq:time,eq:ode,eq:constraints}) to unconstrained minimization of the loss function
\begin{equation} \label{eq:odil}
  \mathcal{L}(\mathbf{x},\theta) =
  \sum\limits_{n=0}^{N-2} \left| \Delta \mathbf{x}^{n+1/2} -  \mathbf{f}^{n+1/2}_\theta \Delta t \right|^2 + \lambda T,
\end{equation}
with respect to the NN parameters $\theta$ and the trajectory $\mathbf{x}$ discretized on a uniform grid of $N$ points in time,
where $\lambda>0$ is a penalization factor.
In \cref{eq:odil}, we used the notation
\begin{gather*}
  \Delta \mathbf{x}^{n+1/2} = \mathbf{x}^{n+1} - \mathbf{x}^{n},
  \\
  \mathbf{f}^{n+1/2}_\theta =
  \mathbf{f}(\mathbf{x}^{n+1/2}, \mathbf{a}_\theta(\mathbf{x}^{n+1/2})),
  \\
  \mathbf{x}^{n+1/2} = \frac{\mathbf{x}^{n+1} + \mathbf{x}^{n}}{2},
  \\
  \Delta t=\frac{\Delta\mathbf{x}\cdot \mathbf{f}_\theta}{\mathbf{f}_\theta\cdot\mathbf{f}_\theta},
  \quad T=(N-1)\Delta t,
\end{gather*}
where $n=0,\dots,N-2$, and the scalar product defined as
$\mathbf{u}\cdot\mathbf{v}
=
\sum_{n=0}^{N-2}\mathbf{u}^{n+1/2}\cdot\mathbf{v}^{n+1/2}$
for vectors $\mathbf{u}$ and $\mathbf{v}$ of length $N$.
This loss function contains the residual of the governing ODE (\cref{eq:ode}) discretized using the midpoint rule on a grid of $N$ points in time.
Unless stated otherwise, the initial and final conditions are imposed exactly by setting
$\mathbf{x}^0=\mathbf{x}_\text{start}$
and $\mathbf{x}^{N-1}=\mathbf{x}_\text{target}$, and the initial guess for the trajectory is a straight line connecting $\mathbf{x}_\text{start}$ and $\mathbf{x}_\text{target}$.
The optimization process relies on AD and the Adam optimization technique.
Further details of the optimization are described in the Supplementary Material, which includes Refs.~\cite{tensorflow2015whitepaper,frostig2018compiling,trottenberg2000multigrid,martin2022,vach2015fast,najafi2004simple,hartl2021,huang2021cem}.

\section{RL approach}
The system is controlled by an agent that determines the appropriate action based on the current state at regular time intervals $\tau > \Delta t$.
The agent's control value, $\mathbf{a}$, remains constant during each time interval.
When it comes to navigation problems, the objective is to reach the target within a close proximity $\delta > 0$ in the shortest possible time.
To represent this objective, a negative reward is assigned after each action.
Furthermore, to aid the system in reaching the target, a reward shaping term is utilized.
This term yields a positive reward when the system's position moves closer to the target following an action~\cite{ng1999}.
Consequently, the reward obtained after the $t$-th action can be expressed as follows:
\begin{equation} \label{e_reward}
  r_t = -\kappa \tau + |\mathbf{x}_t - \mathbf{x}_\text{target}| - |\mathbf{x}_{t+1} - \mathbf{x}_\text{target}|,
\end{equation}
where $\kappa>0$ is denoted as the time penalty coefficient.
We note that the shaping term in \cref{e_reward} does not modify the objective as it is derived from a potential~\cite{ng1999}.
An episode is terminated if the time exceeds a value $T_\text{max}$ or if the system's position has reached the target within the distance $\delta$.
Details about the RL training are listed in the Supplementary Material.

\section{Magnetic swimmers}
\begin{figure}
  \centering
  \includegraphics{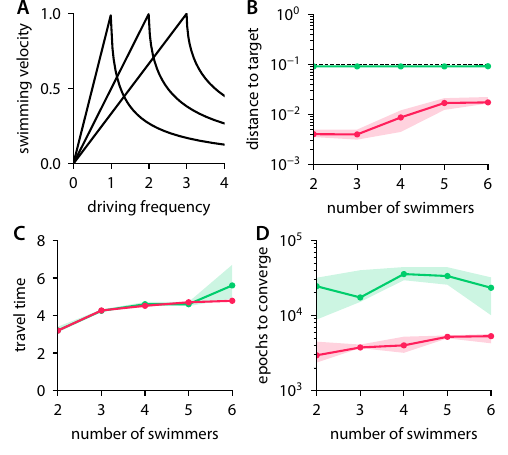}
  \caption{%
    Magnetic swimmers.
    (\figbf{a})~Velocity response of three swimmers depending on the driving frequency.
    (\figbf{b},~\figbf{c},~\figbf{d})~Distance to target, travel time, and number of epochs until convergence versus the number of swimmers for ODIL~\key{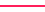} and RL~\key{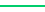}.
    Shades are the $20^\text{th}$ to $80^\text{th}$ percentiles, solid lines are the medians over 10 realizations.
    The dashed line shows the threshold distance to target $\delta$.
  }
  \label{fig:magnswim}
\end{figure}

Artificial bacterial flagella (ABFs) are microswimmers that are propelled through external magnetic fields in viscous flow environments~\cite{zhang2009}.
When these swimmers are immersed in a rotating, uniform magnetic field, they exhibit a magnetic moment, which induces a torque along their primary axis.
The helicoidal shape of ABFs causes the torque to be converted into linear velocity in the direction perpendicular to the plane of rotation of the external magnetic field, denoted as $\mathbf{p}$.
The average velocity magnitude of an ABF is given by~\cite{schamel2013,vach2013}
\begin{equation*}
  v(\omega; b, \omega_c) =
  \begin{cases}
    b \omega, & \text{if} \;\; \omega \leq \omega_c, \\
    b \left(\omega - \sqrt{\omega^2 - \omega_c^2}\right), & \text{otherwise},
  \end{cases}
\end{equation*}
where $\omega$ is the angular frequency of the magnetic field and the parameters $b$ and $\omega_c$ depend on the shape and magnetic properties of the ABF~\cite{schamel2013}.
Despite a uniform magnetic field, it is possible to control multiple ABFs independently if their shapes differ~\cite{amoudruz2022}.
We describe the above system in two dimensions and neglect hydrodynamic and magnetic interactions between swimmers.
The resulting system of ODE reads $\dot{\mathbf{x}}_i = v(\omega; b_i, \omega_{c,i}) \mathbf{p}$, $i=2,\dots,M$, where $M$ is the number of swimmers.
The task consists in guiding all $M$ swimmers from their initial position $\mathbf{x}_{\text{start},i}$ towards a target $\mathbf{x}_\text{target}$ in the least amount of time, by modulating the magnetic field direction $\mathbf{p}$ and angular frequency $\omega$.
The direction of the magnetic field is set to $\mathbf{p} = \mathbf{a}_p / |\mathbf{a}_p|$, with $\mathbf{a}_p = (a_{\theta,1}, a_{\theta,2})$.
the third component of $\mathbf{a}_{\theta}$ encodes the angular frequency $\omega$.
We set $\omega_{c,i} = i$ and $b_i = 1 / i$, $i=1,2,\dots,M$ (\cref{fig:magnswim}).
The target of each ABF is placed at the origin and the starting positions are located on the left side of the unit circle, $\mathbf{x}_{\text{start},i} = (\cos\phi_i, \sin\phi_i)$, with $\phi_i = \pi/2 + i \pi/(M-1)$, $i=1,2,\dots,M$.
See Supplementary Material for details about the ODIL and RL training parameters.

\begin{figure}
  \centering
  \includegraphics{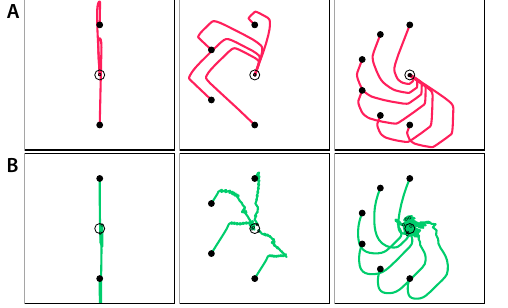}
  \caption{%
    Trajectories of $M=2$, 4, and 6 magnetic swimmers with starting positions~\key{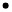} and target~\key{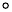}
    using ODIL~\key{110}~(\figbf{a}) and RL~\key{120}~(\figbf{b}).
  }
  \label{fig:magnswim:traj}
\end{figure}

\Cref{fig:magnswim:traj} shows the trajectories of ABFs that follow policies obtained with the two approaches.
Both methods successfully steer all $M$ swimmers for $M\in\{2,4,6\}$.
The trajectories learned with the two methods are different, but we note that this problem has many solutions~\cite{amoudruz2022}.
ODIL converges an order of magnitude faster than RL in terms of number of epochs (\cref{fig:magnswim}).
Both methods seem to find a similar travel time, but the RL approach degrades when the number of swimmers is larger than 5 compared to ODIL.
Furthermore ODIL brings the swimmers closer to the target than RL, which stops when all swimmers reach the target within the distance $\delta = 0.1$.

As the number of ABFs increases, this problem becomes increasingly challenging, especially when the parameters $\omega_{c,i}$ are not well separated.
Furthermore, with a large number of swimmers, hydrodynamic interactions become significant, increasing the difficulty to control ABF swarms.
Addressing the control of such swarms will be the focus of further research.

\section{Transport with vortices}
Here we consider a system of multiple particles transported through a grid of vortices~\cite{ye2012micro,basualdo2022control}.
The task is to bring the particles from their starting positions to individual targets by controlling the intensity of each vortex.
The system consists of $M$ particles passively transported by the vortical flow as $\dot{\mathbf{x}}_i = \mathbf{u}(\mathbf{x}_i)$.
The velocity field is a superposition of $M\times M$ vortices
\begin{equation*}
  \mathbf{u}(\mathbf{x})=
  \sum\limits_{i=1}^{M\times M} \omega_i\,\mathbf{u}_V(\mathbf{x} - \mathbf{c}_i),
\end{equation*}
with intensities $\omega_i\in(-1,1)$ and centers $\mathbf{c}_i$ forming a uniform grid in $[0.5, M - 0.5]\times[0.5, M - 0.5]$, and each vortex produces a flow proportional to $\mathbf{u}_V(x, y) = e^{-(x^2+y^2)/0.72}(-y, x)$.
The task is to transport the particles from their starting positions $\mathbf{x}_{\text{start},i} = (0.5 + i, 0)$ to targets $\mathbf{x}_{\text{target},i} = (0.5 + i, M)$ in the minimal time $T$ by controlling the vortex intensities $\omega_i$ as a function of the current positions of all beads.
See Supplementary Material for details about the ODIL and RL parameters.

\begin{figure}
  \centering
  \includegraphics{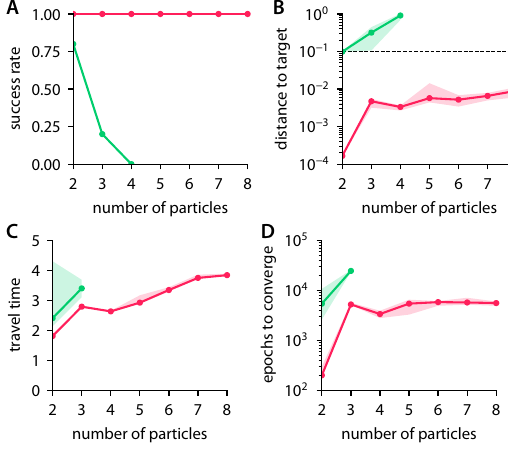}
  \caption{%
    Manipulation by vortices.
    Performance of ODIL~\key{110} and RL~\key{120} depending on the number of particles:
    success rate (\figbf{a}), distance to target~(\figbf{b}),
    number of epochs until convergence~(\figbf{c}), and travel time~(\figbf{d}).
    Shades are the $20^\text{th}$ to $80^\text{th}$ percentiles, solid lines are the medians over 10 realizations.
    The dashed line shows the threshold distance to target $\delta$.
  }
  \label{fig:vortmanip}
\end{figure}

\Cref{fig:vortmanip} compares the control policies of ODIL and RL for $M=2,\;3,\dots,\;8$ swimmers.
Example trajectories obtained with ODIL are shown in \cref{fig:vortmanip:odil}.
See Supplementary Material for trajectories obtained with RL and for other values of $M$.
For $M=2$, both methods find similar trajectories in all realizations.
For $M=3$, only 30\% of realizations of RL lead to a valid path to the targets, and the successful trajectories appear more erratic than those found by ODIL.
For larger values $M\geq 4$, the RL does not produce any valid policy, while ODIL successfully solves the problem.
ODIL tends to gather the particles in a central corridor before dispatching each particle to their respective targets.
Since the dimension of the action space grows quadratically with $M$, the use of gradients by ODIL becomes advantageous over RL for finding valid paths.

\begin{figure}
  \centering
  \includegraphics{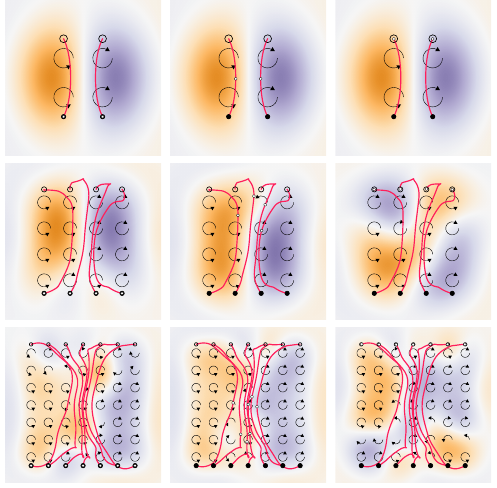}
  \caption{%
    Manipulation by vortices using ODIL~\key{110} for $M=2$, 4, and 7.
    Each row corresponds to one trajectory at times $t/T=0$, 0.5, and 1.
    Vorticity field with arrows showing the intensity and orientation of vortices.
  }
  \label{fig:vortmanip:odil}
\end{figure}

\section{Computational cost}
We compare the cost of ODIL against RL through two metrics.
The first metric is the number of forward and backward passes through the NN until convergence, and is relevant when solving the dynamics of the system is relatively cheap.
The second metric is wall time until convergence, which reflects both the back-propagation through the NN and solving the dynamics of the environment.
The RL approach uses a replay memory buffer to store past experiences, used in mini-batches of size $B$ to estimate the off-policy objective loss.
Each training step thus corresponds to $B+1$ forward passes and one backward pass through the NN.
In this study we used one training step per experience and $B=256$.
The ODIL approach, on the other hand, requires one forward and one backward passes per discretization point along one trajectory.
In all cases, we define that the training is converged when the travel time is within $1\%$ of the best value.

\begin{table}
  \centering
  \begin{tabular}{l|c|c|c|c}
    \textbf{Case}      & \multicolumn{2}{c}{\textbf{Policy evaluations}} & \multicolumn{2}{c}{\textbf{Wall time (s)}} \\
                       & ODIL                     & RL              & ODIL                    & RL               \\ \hline
    ABFs, $M=3$        & \textbf{\num{1.29e+06}}  & \num{5.47e+08}  & \textbf{\num{1.68e+02}} & \num{1.55e+04}   \\
    ABFs, $M=5$        & \textbf{\num{1.29e+06}}  & \num{8.44e+08}  & \textbf{\num{1.75e+02}} & \num{2.45e+04}   \\
    vortices, $M=2$    & \textbf{\num{1.29e+06}}  & \num{1.35e+08}  & \textbf{\num{9.90e+01}} & \num{4.72e+03}   \\
    vortices, $M=3$    & \textbf{\num{1.29e+06}}  & \num{6.18e+08}  & \textbf{\num{1.04e+02}} & \num{2.54e+04}   \\
    vortices, $M=8$    & \textbf{\num{1.29e+06}}  & N/A             & \textbf{\num{1.44e+02}} & N/A
  \end{tabular}
  \caption{
    Number of policy evaluations and total wall time for training the policy with ODIL and RL (median over 10 random seeds).
  } \label{tab:cost}
\end{table}

We report these metrics for the two cases detailed in this letter with different values of $M$.
Additional examples in the Supplementary Material show similar trends.
It is clear from \cref{tab:cost} that for all benchmark problems, the computational cost is orders of magnitude lower with ODIL compared to RL.
While the number of training epochs was similar for both methods, the RL approach requires to evaluate the policy over mini-batches at each policy update, explaining the large cost associated with this approach.
This resulted in runtimes larger by factors of 10-100 for the RL approach over ODIL.
We remark that the RL method does not always converge to a satisfactory solution within a reasonable time, and the numbers reported in \cref{tab:cost} are computed among the successful trials, when available.
We expect that ODIL remains advantageous even for more costly dynamical systems.
Indeed, in addition to the faster convergence of ODIL over RL, ODIL is more robust in finding satisfactory solutions at higher state/action spaces.

We note that RL and ODIL operate under fundamentally different assumptions about system dynamics.
ODIL, a model-based approach, requires a differentiable ODE model of system dynamics, whereas model-free RL relies on substantial environment interaction to converge, which may explain their performance differences.
In cases of incomplete system knowledge, ODIL could potentially integrate a data-driven model of unknown dynamics, which will be explored in future work.
Finally, while this work uses a single initial condition, the Supplementary Material includes two examples demonstrating closed-loop control with multiple initial conditions, enhancing resilience to external noise.

\section{Summary}
We have presented the ODIL method to solve closed-loop control and navigation problems described by ODEs.
The formulation of the method relies on the optimization of a loss function that combines the residuals of the governing equations, the initial and target positions of the system, and the objective.
We perform the optimization with standard machine learning tools that allows us to seamlessly describe the control policy with NNs.
We have tested the new method on a variety of navigation problems and compared against a state of the art RL algorithm.
In all cases, ODIL has a lower computational cost than RL while producing policies with similar or better performance.
Furthermore, we have shown that RL fails to reach the target when the action space is high-dimensional, while ODIL reliably solves the problem in such cases.

ODIL is well suited for problems with known targets due to its implicit formulation and the use of gradients.
On the other hand, RL requires reward shaping and sometimes extensive tuning to work correctly.
Future applications of ODIL can consider more complex dynamical systems such as those described by partial differential equations.

\section{Acknowledgments}
Authors acknowledge the use of computing resources
from the Swiss National Supercomputing Centre (CSCS) under project s1160.

\bibliographystyle{unsrt}
\bibliography{bibliography}

\clearpage

\renewcommand{\thesection}{S\arabic{section}}
\renewcommand{\thesubsection}{S\arabic{section}.\arabic{subsection}}
\renewcommand{\thesubsubsection}{S\arabic{section}.\arabic{subsection}.\arabic{subsubsection}}
\renewcommand{\thefigure}{S\arabic{figure}}
\renewcommand{\thetable}{S\arabic{table}}
\renewcommand{\theequation}{S\arabic{equation}}
\makeatletter
\renewcommand{\theHsection}{S\arabic{section}}
\renewcommand{\theHsubsection}{S\arabic{section}.\arabic{subsection}}
\renewcommand{\theHtable}{S\arabic{table}}
\renewcommand{\theHfigure}{S\arabic{figure}}
\renewcommand{\theHequation}{S\arabic{equation}}
\makeatother
\setcounter{figure}{0}
\setcounter{table}{0}
\setcounter{equation}{0}

\setcounter{section}{0}

\section*{Supplementary Information}

\section{Details about Optimization Process}

We solve the optimization with a gradient-based optimizer Adam~\cite{kingma2014adam}.
Gradients are evaluated through automatic differentiation implemented in Tensorflow~\cite{tensorflow2015whitepaper}
or JAX~\cite{frostig2018compiling}.
The loss function contains both the residuals of
the governing equations and the objective to be minimized.
To solve the constrained minimization problem,
we perform several rounds of unconstrained optimization
decreasing the penalization factor $\lambda$ exponentially for each round.
Throughout the paper, one epoch refers to one iteration of the optimizer.
While ODIL produces an approximate solution to the ODEs as part of optimization,
the results in the paper, such as the trajectories, measurements of the achieved
distance and travel time, come from evaluations of the policy on the forward problem solved with the explicit RK2 method
with a smaller time step of $\Delta t/4$,
on a grid of $4(N-1)+1$ points in time.

To accelerate the convergence, we use the multigrid
decomposition technique to represent the unknown discretized fields that are optimized with ODIL~\cite{karnakov2023flow}.
Consider a hierarchy of successively coarser grids of size
$N_i = N / 2^{i-1}$ cells for $i=1,\dots,L$,
where $L$ is the total number of levels.
Define the multigrid decomposition operator as
\begin{equation}
M_L(u_1, \dots, u_L) = u_1 + T_1 u_2 + \dots +  T_1 T_2 \dots T_{L-1} u_{L},
\end{equation}
where each $u_i$ is a field on grid $N_i$,
and each $T_i$ is an interpolation operator from grid $N_{i+1}$ to the finer grid~$N_i$.
The multigrid decomposition of a discrete field $u$ on a grid of size $N$ reads
\begin{equation}
u = M_L(u_1, \dots, u_L).
\end{equation}
Note that this representation is over-parameterized and therefore not unique.
The total number of scalar parameters increases from $N^d$ of the original field
$u$ to $N_1^d + \dots + N_L^d$ for the representation $u_1,\dots,u_L$.
We define the interpolation operators~$T_i$ using linear
interpolation~\cite{trottenberg2000multigrid}
for node-based discretizations.
This technique addresses the issue of locality of gradient-based optimizers
by extending the domain of dependence of each scalar parameter
so that information can propagate through the grid faster.

\section{Evaluation of the ODIL policy}

To evaluate the performance of ODIL, we proceed as follows.
The policy is first trained as described in the main text, and we store the weights of the NN.
We then evaluate the obtained policy by integrating the dynamical system with an explicit RK2 time integrator.
The time step is chosen as $\Delta t/4$, where $\Delta t$ is the time step that was used in ODIL.
Decreasing this time step did not change the results.

\section{Reinforcement Learning}

The agent is trained with the off-policy actor-critic algorithm V-RACER~\cite{novati2019}, implemented in the software Korali~\cite{martin2022}.
A NN maps the observables of the system to an approximation of the value function and the parameters that describe the distribution of the actions (mean and variance for each variable).
While training, each action is sampled from a truncated Gaussian with the parameters given by the NN.
During the testing phase, the agent takes greedy actions and the policy is thus deterministic.
The training aims to minimize the difference between the value function coming from the NN and that estimated from $B$ experiences sampled from the replay memory.
The sampling of these experiences follows the remember and forget experience replay (REFER) approach as described in~\cite{novati2019}.

\section{Magnetic swimmers}

The parameters of the swimmers, $\omega_{c,i} = i$, $b_i = 1/i$, were chosen arbitrarily.
Nevertheless, these parameters can be realized experimentally.
The slope $b$ of the velocity response depends solely on the shape of the swimmer, and can be adapted by changing the chirality of the swimmer, spanning a large range of values for $b$~\cite{vach2015fast}.
The range of values observed for random shapes spans 2 orders of magnitude~\cite{vach2015fast}, and thus it is realistic to obtain the values of $b_i = 1/i$ for $M \leq 6$.
The critical angular frequency $\omega_c$, for a given swimmer shape and a given external field magnitude, depends only on the magnetic moment of the swimmer.
This value can be tuned by changing the amount of magnetic material within the swimmer.

In the ODIL setup,  the grid size is $N=129$ points in time.
The policy is a fully connected NN of size $128\times 128$
with inputs $(\mathbf{x}_1, \dots, \mathbf{x}_M)$ and outputs $(\omega,\mathbf{p})$.
The output $\mathbf{p}$ is normalized so that $|\mathbf{p}|=1$,
and a $\tanh$ function maps $\omega$ into $(0.9\,\omega_{c,1}, 1.1\,\omega_{c,M})$.
The starting positions and targets are imposed exactly.
The optimization consists of 10 rounds of 1000 iterations,
with the penalization factor $\lambda=0.02$ and learning rate $\eta=0.005$ in the first five rounds that are multiplied by 0.25 before each of the subsequent five rounds.
To enable exploration, the output $\mathbf{p}$ of the policy network is perturbed by noise sampled from the normal distribution
with a standard deviation of $1/\sqrt{N}$.
The perturbation remains constant, i.e. the noise is sampled once, and only applies to the first round of optimization.

In the RL approach, an episode is successful when every swimmer is within a distance $\delta$ from its target.
The reward is set to:
\begin{equation}
  r_t = -\kappa \tau +
  \sum\limits_{i=1}^M |\mathbf{x}_i(t-\tau) - \mathbf{x}_{\text{target},i}| -
  \sum\limits_{i=1}^M |\mathbf{x}_i(t) - \mathbf{x}_{\text{target},i}|.
\end{equation}
We choose a time limit $T_\text{max} = 20$, time penalty coefficient $\kappa=10$, time between actions $\tau = 0.1$, distance to target $\delta=0.1$, and train for 50\,000 episodes on 10 random seeds.
The integration time step in RL was set to $\Delta t = 0.01$.

\begin{figure}[h]
  \centering
  \includegraphics{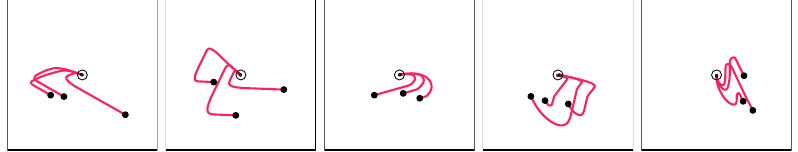}
  \caption{%
    Trajectories of 3 magnetic swimmers starting from random positions, following the ODIL policy.
  }
  \label{fig:magnswim:rand:odil}
\end{figure}

\Cref{fig:magnswim:rand:odil} shows trajectories of magnetic swimmers that start at random initial positions uniformly sampled in the domain $[-1, 1] \times [-1, 0]$.
In each case, the policy is trained as described above and the trajectories are computed by solving the ODE with a Runge-Kutta integrator of second order.

\section{Particle transport with vortices}

In the ODIL setup, we use $N=129$ grid points along the time direction.
The policy is a fully connected NN of size $128\times 128$
with inputs $(\mathbf{x}_1, \dots, \mathbf{x}_M)$ and outputs $(\omega_1, \dots, \omega_M)$.
The output layer has a $\tanh$ activation to map the outputs into $(-1, 1)$.
The starting positions and targets are imposed exactly.
The optimization consists of 10 rounds of 1000 iterations,
with the penalization factor $\lambda=0.1$ and learning rate $\eta=0.005$ in the first five rounds that are multiplied by 0.5 before each of the subsequent five rounds.

In the RL setup, the reward is
\begin{equation}
  r_t = -\kappa \tau +
  \sum\limits_{i=1}^M |\mathbf{x}_i(t-\tau) - \mathbf{x}_{\text{target},i}|
  - \sum\limits_{i=1}^M |\mathbf{x}_i(t) - \mathbf{x}_{\text{target},i}|,
\end{equation}
the action consists of the vortex intensities $\omega_i$,
an episode is successful when every particle is within a distance $\delta=0.1$ from its target.
We use the time limit $T_\text{max}=10$, time penalty coefficient $\kappa=10$, time between actions $\tau=0.1$ and distance to target $\delta=0.1$, and train for 50\,000 episodes.
The integration time step in RL was set to $\Delta t = 0.005$.

\begin{figure}
  \centering
  \includegraphics{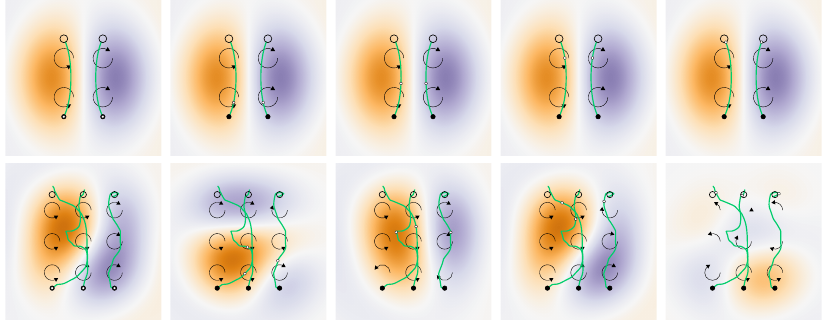}
  \caption{%
    Manipulation by vortices using RL~\key{120} for $M=2$ and 3.
    Each row corresponds to one trajectory at times $t/T=0$, 0.25, 0.5, 0.75, and 1.
    Vorticity field with arrows showing the intensity and orientation of vortices.
  }
  \label{fig:vortmanip:s_rl}
\end{figure}

\begin{figure}
  \centering
  \includegraphics{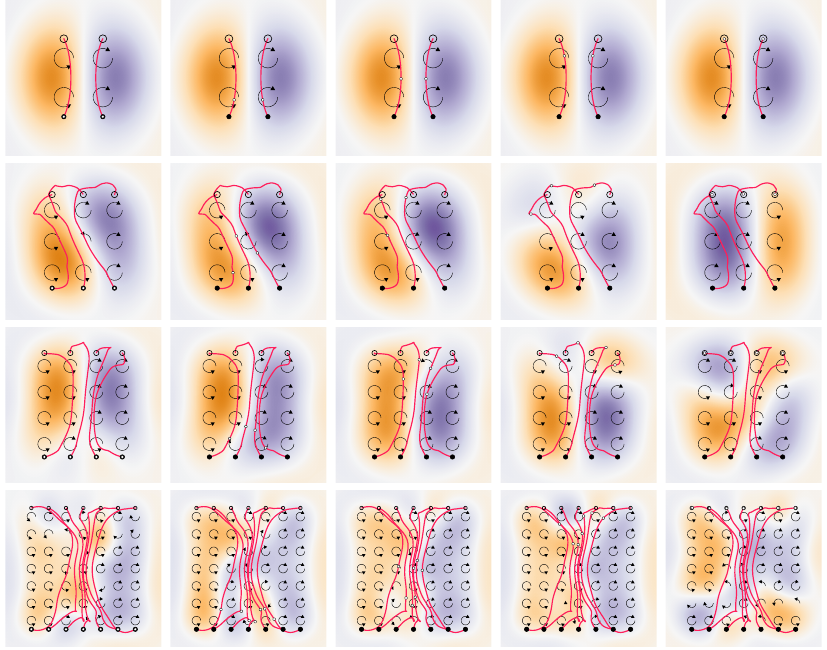}
  \caption{%
    Manipulation by vortices using ODIL~\key{110} for $M=2$, 3, 4, and 7.
    Each row corresponds to one trajectory at times $t/T=0$, 0.25, 0.5, 0.75, and 1.
    Vorticity field with arrows showing the intensity and orientation of vortices.
  }
  \label{fig:vortmanip:s_odil}
\end{figure}

Trajectories obtained with the RL and ODIL policies are shown on \cref{fig:vortmanip:s_rl,fig:vortmanip:s_odil}, respectively.

\section{Other examples}

\subsection{Brachistochrone}
\label{s_brach}

The brachistochrone problem involves finding the curve along which a particle will slide from one point to another in the shortest time under the influence of gravity, assuming no friction. Mathematically, this can be described by a particle moving along a parametric curve $\mathbf{x}(t) = (x(t), y(t))$
 while being accelerated by gravity $g$ without friction.
The problem consists in finding a curve that minimizes the travel time $T$ from a starting position $\mathbf{x}(0) = \mathbf{x}_\text{start}$ to a target position $\mathbf{x}(T) = \mathbf{x}_\text{target}$ with zero initial velocity~$\dot{\mathbf{x}}(0)=\mathbf{0}$.
The motion of the particle is described by the ODE
\begin{equation} \label{e_brach_ode}
  \begin{aligned}
    \dot{x} &= v \cos\phi, \\
    \dot{y} &= v \sin\phi, \\
    \dot{v} &= -g \sin\phi,
  \end{aligned}
\end{equation}
where $v$ is the velocity magnitude of the particle and $\phi$ is the angle between the tangent $\dot{\mathbf{x}}$ and the $x$-axis.
We set $g=1$, $\mathbf{x}_\text{start} = (0, 0)$ and $\mathbf{x}_\text{target} = (3, 0)$.

In the ODIL setup, the grid size is $N=129$ points in time.
The policy~$\mathbf{a}_\theta(\mathbf{x})$ is a fully connected NN of size $128\times 128$
with inputs $(x, y)$ and output $\phi$.
The output layer has a $\tanh$ activation to map the outputs into $(-\pi/2 ,\pi/2)$.
The known initial and final conditions are imposed exactly, and the final condition for velocity $v$ remains unspecified.
The optimization consists of 10 rounds of 100 iterations,
with the penalization factor $\lambda=0.1$ and learning rate $\eta=0.005$ in the first five rounds that are multiplied by 0.5 before each of the subsequent five rounds.

In the RL setup, we formulate the task as a decision process, where the agent travels on the curve and controls its direction parameterized by $\phi \in [-\pi/2, \pi/2]$.
The state variable is the relative position of the agent to the target, $\mathbf{x} - \mathbf{x}_\text{target}$.
An episode ends when either the simulation time $t$ exceeds a limit $T_\text{max}$, or when the agent's position along the $x$-axis exceeds that of the target.
The reward comprises two terms, $r_t = c_t + d_t$, corresponding to reaching the target and doing so in the shortest possible time.
Reaching the target is encoded by adding a term at the end of the episode:
\begin{equation} \label{e_brach_reward}
  c_t =
  \begin{cases}
    -k_x |x-x_\text{target}|, & \text{if}\;\; t \geq T_\text{max}, \\
    k_y \exp \left( - (y - y_\text{target})^2 / (2 \sigma^2) \right), & \text{if} \;\; x \geq x_\text{target}, \\
    0 & \text{otherwise},
  \end{cases}
\end{equation}
where $k_x$, $k_y$ and $\sigma$ are positive constants.
To penalize long trajectories, the additional term $d_t = -\tau / T_\text{max}$ is added at every step.
Here we used $k_x=1$, $k_y=500$, $\sigma=0.2$, $\tau = 0.02$, $\Delta t = 2\times 10^{-4}$, and $T_\text{max}=10$.
We remark that the RL approach failed with the reward formulation that was used in the rest of this study, and thus we opted for \cref{e_brach_reward} with extensive reward shaping.

\begin{figure}
  \centering
  \includegraphics{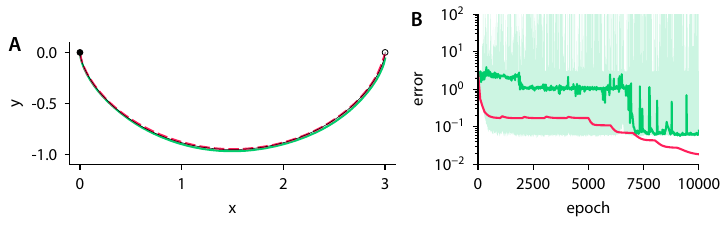}

  \caption{%
    Brachistochrone.
    (\figbf{a})~Trajectory from ODIL~\key{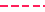} and RL~\key{120} compared to the exact solution~\key{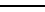}.
    The starting position is $(0, 0)$~\key{001} and target is $(3,0)$~\key{001open}.
    (\figbf{b})~Training history showing the error compared to the exact trajectory for ODIL~\key{110} and RL~\key{120}.
    Shades are the $20^\text{th}$ to $80^\text{th}$ percentiles, solid lines are the medians over 10 realizations.
  }
  \label{f_brach}
\end{figure}

\Cref{f_brach} shows the trajectories obtained with ODIL and RL, along with the exact trajectory found analytically.
We define the error as
\begin{equation} \label{e_brach_error}
  \mathrm{Error} =
  \Big(\tfrac{1}{x_\text{target}}\int_0^{x_\text{target}}
    \big(y(x) - y_\text{exact}(x)\big)^2dx
    + \big|\mathbf{x}(T) - \mathbf{x}_\text{target}\big|^2
  \Big)^{1/2}
\end{equation}
to measure the difference with the exact trajectory and the proximity of the target.
Both approaches find the optimal trajectory and typically take about 7\,000 epochs to achieve an error of 0.05.
The results include 10 realizations with different random seeds.
RL reaches an error below 0.1 in only 60\% of the cases, while the results of ODIL do not significantly depend on the random seed.
We remark that designing a successful RL setup took several attempts with various definitions of the reward and values of the hyper-parameters.
ODIL is more robust in this case and follows the general approach without extensive parameter tuning.

\subsection{Path planning in a 2D background flow}
\label{s_pathplan}

We consider the problem of path planning for a swimmer at position $\mathbf{x}(t)$ in a two-dimensional domain.
The swimmer propels itself at a constant swimming velocity $U>0$ relative to a steady background flow $\mathbf{u}(\mathbf{x})$ without inertia
\begin{equation} \label{e_pathplan_ode}
  \begin{aligned}
    \dot{\mathbf{x}} &= \mathbf{f},\\
    \mathbf{f} &= \mathbf{u} (\mathbf{x}) + U \mathbf{p},
  \end{aligned}
\end{equation}
where the unit vector $\mathbf{p}$ is the instantaneous swimming direction.
The task is to find $\mathbf{p}$ that brings the swimmer
from a starting position $\mathbf{x}_\text{start}$
to a target position $\mathbf{x}_\text{target}$ in the minimal time $T>0$.
The action sets the swimming direction as $\mathbf{p} =\mathbf{p}_\theta = \mathbf{a}_\theta / |\mathbf{a}_\theta|$.
We solve this problem using ODIL and RL in two cases with different choices of the background flow: simple shear and vortical flow.

The ODIL setup follows the method explained in the manuscript except that the target conditions are imposed with a penalty term.
The loss function includes the residuals of the governing equations~\cref{e_pathplan_ode},
a term to impose the target condition, and the objective
\begin{equation} \label{e_pathplan_odil_loss}
  \mathcal{L}(\mathbf{x},\theta) =
  \sum\limits_{n=0}^{N-2} \left| \Delta \mathbf{x}^{n+1/2}
  -\mathbf{f}^{n+1/2}_\theta \Delta t \right|^2
  + k_\text{imp}|\mathbf{x}^{N-1}-\mathbf{x}_\text{target}|^2
  + \lambda T,
\end{equation}
where $k_\text{imp}=10$, while the initial conditions are imposed exactly.
We have found that this penalty formulation for the target conditions
tends to be more robust in finding the global minimum.
The initial guess for the trajectory is $\mathbf{x}=\mathbf{x}_\text{start}$.

\begin{figure}
  \centering
  \includegraphics{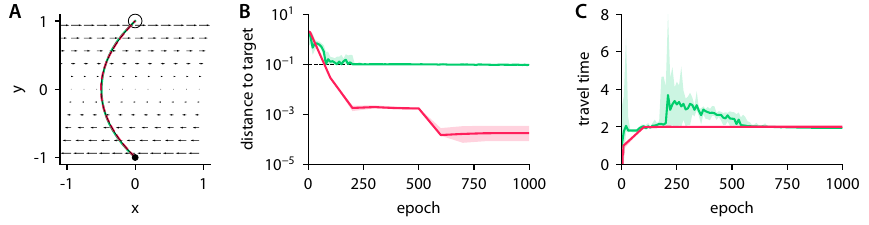}

  \caption{
    Path planning in simple shear using ODIL and RL.
    (\figbf{a})~Trajectories from ODIL~\key{210}, RL~\key{120}, and exact solution~\key{100}
    with velocity of the background flow (arrows), starting position $(0,-1)$~\key{001}, and target $(0,1)$~\key{001open}.
    (\figbf{b},~\figbf{c})~Training history showing
    the distance from the nearest point on the trajectory to the target,
    and the time to reach that nearest point for ODIL~\key{110} and RL~\key{120}.
    Shades are the $20^\text{th}$ to $80^\text{th}$ percentiles, solid lines are the medians over 10 realizations.
  }
  \label{f_pathplan_shear}
\end{figure}

In the first case, the background flow is characterized by a simple shear velocity field $\mathbf{u}(x, y) = (y, 0)$.
We set the swimming velocity to~$U=1$, starting position to $\mathbf{x}_\text{start} = (0,-1)$, and target to $\mathbf{x}_\text{target} = (0,1)$.
In the ODIL setup, the grid size is $N=129$ points in time, the optimization consists of 10 rounds of 100 iterations,
with the penalization factor $\lambda=0.01$ and learning rate $\eta=0.005$ in the first five rounds
that are multiplied by 0.1 before each of the subsequent five rounds.
In the RL setup, we use the time limit $T_\text{max} = 10$, time between actions $\tau = 0.1$, and distance tolerance $\delta=0.1$ and train for 10\,000 episodes.
The integration time step in RL was set to $\Delta t = 0.01$.
The optimal trajectory, found analytically~\cite{liebchen2019},
is a parabola which corresponds to the constant policy $\mathbf{p}=(0, 1)$ and the optimal travel time of $T=2$.
The trajectories obtained with ODIL and RL are shown in~\cref{f_pathplan_shear}.
Both methods find the optimal trajectory and travel time.
RL takes about 600 epochs to reach the target and converge to the optimal travel time, while ODIL only takes 100 epochs.
The minimal distance to target in \cref{f_pathplan_shear} reached by RL
equilibrates at the specified distance tolerance~$\delta$ while ODIL reaches smaller values up to $10^{-4}$.
The results include 10 realizations with different random seeds,
and both methods successfully find the optimal trajectory in all realizations without significant scatter.

\begin{figure}
  \centering
  \includegraphics{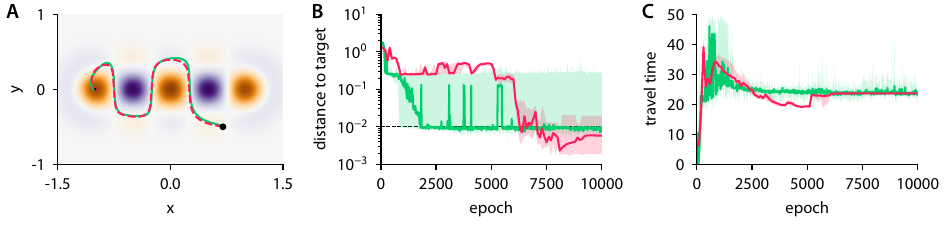}

  \caption{%
    Path planning in vortical flow using ODIL and RL.
    (\figbf{a})~Trajectories from ODIL~\key{210} and RL~\key{120}
    with vorticity of the background flow, starting position $(0.7,-0.5)$~\key{001}, and target $(0,1)$~\key{001open}.
    (\figbf{b},~\figbf{c})~Training history showing
    the distance from the nearest point on the trajectory to the target,
    and the time to reach that nearest point for ODIL~\key{110} and RL~\key{120}.
    Shades are the 20\% and 80\% percentiles, solid lines are the medians over 10 realizations.
  }
  \label{f_pathplan_gears}
\end{figure}

In the second case, the background flow is a superposition of five vortices placed
on the $x$-axis and a uniform flow.
\begin{equation}
  \mathbf{u}(x, y) =
  \mathbf{u}_V(x + 1, y) - \mathbf{u}_V(x + 0.5, y) + \mathbf{u}_V(x, y) - \mathbf{u}_V(x - 0.5, y) + \mathbf{u}_V(x - 1, y)
  + \mathbf{u}_0,
\end{equation}
where each vortex is
\begin{equation}
  \mathbf{u}_V(x, y) = 2\, e^{-({x^2+y^2})/0.08} (-y, x)
\end{equation}
and the uniform flow is~$\mathbf{u}_0=(0.2,0)$.
We set the swimming velocity to~$U=0.22$, starting position to $\mathbf{x}_\text{start} = (0.7,-0.5)$, and target to $\mathbf{x}_\text{target} = (-1,0)$.
In the RL setup we use $T_\text{max} = 50$, time between actions $\Delta t = 0.2$, and distance tolerance $\delta=0.01$ and train for 20\,000 episodes.
In the ODIL setup, the grid size is $N=129$ points in time, the optimization consists of 10 rounds of 1\,000 iterations,
with the penalization factor $\lambda=0.01$ and learning rate $\eta=0.001$ in the first five rounds
that are multiplied by 0.1 before each of the subsequent five rounds.
The trajectories obtained with ODIL and RL are shown in~\cref{f_pathplan_gears}.
Both methods follow the same strategy, exploiting  vortices to reach the target faster,
typically take about 7\,000 epochs to reach the target withing the specified distance tolerance~$\delta$,
and achieve a travel time of about $T=24$.
The results include 10 realizations with different random seeds.
RL is less robust than in the previous case, as some realizations do not approach the target within a distance of 0.4, i.e. fail to pass all five vortices.
ODIL approaches the target within a distance of 0.07 or less in all cases.

\subsection{Path planning with multiple trajectories and stochastic dynamics}
\label{s_pathplan_noise}

\begin{figure}
  \centering
  \includegraphics{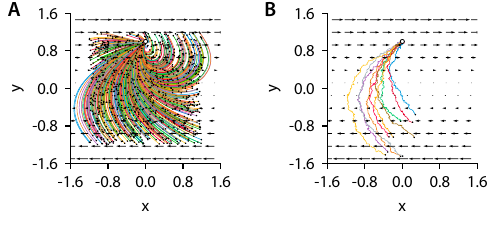}

  \caption{
    Path planning in simple shear with multiple trajectories using ODIL.
    (\figbf{a})~Trajectories from 1000 starting positions~\key{001}
    with a single target $(0,1)$~\key{001open} and common policy.
    (\figbf{b})~Trajectories with added noise from 10 starting positions
    using policy from $\figbf{a}$.
  }
  \label{f_pathplan_noise}
\end{figure}

Here we solve the path planning problem with ODIL for multiple trajectories
at once with the same policy.
We then apply the resulting policy to the same system with added noise on its right hand side to demonstrate its robustness.
We extend the previous case of~\cref{s_pathplan}
by including multiple trajectories in the optimization problem.
The trajectories differ by the choice of the starting position.
We sample $S=1000$ starting positions $\mathbf{x}_{s,\text{start}}$ from
a uniform distribution over the rectangle $\mathbf{x}\in [-1.2,1.2]\times [-0.2,2.2]$
and impose the initial condition $\mathbf{x}_s^0 = \mathbf{x}_{s,\text{start}}$ exactly,
where $\mathbf{x}^n_s$ the swimmer position on trajectory $s=1,\dots,S$ at time step $n=0,\dots,N-1$.
The loss function includes the residuals of the governing equations~\cref{e_pathplan_ode},
terms to impose the target condition, and objectives for each trajectory
\begin{equation} \label{e_pathplan_noise_odil_loss}
  \mathcal{L}(\mathbf{x},\theta) =
  \frac{1}{S}\sum\limits_{s=1}^{S} \left[
  \sum\limits_{n=0}^{N-2} \left| \Delta \mathbf{x}^{n+1/2}_s
  -\mathbf{f}^{n+1/2}_{\theta,s} \Delta t \right|^2
  + k_\text{imp}|\mathbf{x}^{N-1}_s-\mathbf{x}_\text{target}|^2
  + \lambda T_s\right],
\end{equation}
where $k_\text{imp}=10$.
The initial guess for each trajectory is $\mathbf{x}_s=\mathbf{x}_{s,\text{start}}$.
The background flow is simple shear $\mathbf{u}(x, y) = (y, 0)$,
the swimming velocity is~$U=1$, and the target is $\mathbf{x}_\text{target} = (0,1)$.
The grid size is $N=129$ points in time.
The optimization consists of 10 rounds of 1000 iterations,
starting with the penalization factor $\lambda=0.01$ and learning rate $\eta=0.005$ in the first five rounds
that are multiplied by 0.1 before each of the subsequent five rounds.
The obtained trajectories are shown in~\cref{f_pathplan_noise}.
The optimization takes about 50~s on a GPU Nvidia A100.

To demonstrate that the obtained policy $\mathbf{p}_\theta$ is robust
to noise, we evaluate the policy on the following stochastic process
\begin{equation}
  \mathbf{x}^{n+1} = \mathbf{x}^{n}
    + \Delta t \big(\mathbf{u}(\mathbf{x}^n) + U\mathbf{p}_\theta(\mathbf{x}^n)\big)
    + \sqrt{D\Delta t} \mathbf{\xi}^n,
\end{equation}
where $D=0.01$, $\Delta t=0.0058$, and $\mathbf{\xi}^n$ are independent standard normal random variables.
The integration continues until $|\mathbf{x} - \mathbf{x}_\text{target}| < 0.02$.
The obtained trajectories are shown in~\cref{f_pathplan_noise}
for 10 samples of the starting position. All trajectories reach the target.

\subsection{Three-bead swimmer}
\label{s_threebeads}

We consider the problem of finding the swimming strategy of a three-bead swimmer~\cite{najafi2004simple,hartl2021}.
The swimmer consists of three beads of radius $R$ located at positions $x_1<x_2<x_3$ on a straight line, linked with two arms of length $L_1 = x_2 - x_1$ and $L_2 = x_3 - x_2$, and evolves in a viscous fluid.
In the limit of no inertia and $R/L_1,R/L_2 \ll 1$, the beads move according to
\begin{equation} \label{e_3beads_ode}
  \begin{aligned}
    \dot{x}_1 &= \frac{f_1}{6\pi R} + \frac{f_2}{4\pi L_1} + \frac{f_3}{4\pi (L_1+L_2)},  \\
    \dot{x}_2 &= \frac{f_1}{4\pi L_1} + \frac{f_2}{6\pi R} + \frac{f_3}{4\pi L_2}, \\
    \dot{x}_3 &= \frac{f_1}{4\pi (L_1+L_2)} + \frac{f_2}{4\pi L_2} + \frac{f_3}{6\pi R},
  \end{aligned}
\end{equation}
where we chose units where the viscosity is 1 and with the following forces applied to each bead
\begin{align*}
  f_1 &= F_1 + f_R(L_1), \\
  f_2 &= -f_1 - f_3, \\
  f_3 &= F_3 - f_R(L_2),
\end{align*}
where $F_1$ and $F_3$ are control variables bounded as $|F_1|\leq1$ and $|F_3|\leq1$.
The spring force $f_R$ is given by
\begin{equation}
  f_R(L) =
  \begin{cases}
    (L - L_\text{min})/R, & \text{if} \;\; L < L_\text{min}, \\
    (L - L_\text{max})/R, & \text{if} \;\; L > L_\text{max}, \\
    0,                    & \text{otherwise},
  \end{cases}
\end{equation}
where $L_\text{min} = 0.7$ and $L_\text{max} = 1.3$.
We set the bead radius to $R = 0.1$ and initial positions to $x_1(0) = -1$, $x_2(0) = 0$, $x_3(0) = 1$.
The task is to find a control policy that propels the swimmer in the positive direction as fast as possible.
This case differs from the original formulation since the target position is unknown.
Instead, we fix the travel time at $T=30$ and maximize the position of the center of mass $x_c = (x_1 + x_2 + x_3) / 3$ at time $T$.
The optimization problem therefore reads
\begin{equation} \label{e_3beads_opt}
  \begin{aligned}
    \text{maximize }&x_c(T)
    \\
    \text{subject to }&\dot{\mathbf{x}} = \mathbf{f}(\mathbf{x}, \mathbf{a}(\mathbf{x})),
    \quad t\in(0,T),
    \\
    &\mathbf{x}(0) = \mathbf{x}_\text{start},
  \end{aligned}
\end{equation}
where $\mathbf{x}=(x_1, x_2, x_3)$, $\mathbf{a}=(F_1, F_3)$, $\mathbf{x}_\text{start}=(-1, 0, 1)$,
and $\mathbf{f}$ is the right-hand side of equations~\cref{e_3beads_ode}.

In the ODIL setup, we reduce the problem to unconstrained minimization of the loss function
\begin{equation} \label{e_3beads_odil}
  \mathcal{L}(\mathbf{x},\theta) =
  \frac{1}{N}\sum\limits_{n=0}^{N-2} \left| \frac{\Delta \mathbf{x}^{n+1/2}}{\Delta t} -  \mathbf{f}^{n+1/2}_\theta\right|^2
  - \lambda x_c^{N},
\end{equation}
where $\lambda>0$ is a penalization factor,
$\mathbf{f}^{n+1/2}_\theta = \mathbf{f}(\mathbf{x}^{n+1/2}, \mathbf{a}_\theta(\mathbf{x}^{n+1/2}))$.
The policy~$\mathbf{a}_\theta(\mathbf{x})$ is a fully connected NN of size $128\times 128$
with inputs $(L_1, L_2)$ and outputs $(F_1, F_3)$.
The output layer has a $\tanh$ activation to map the outputs into $(-1 ,1)$.
The grid consists of $N=1025$ points in time.
The optimization consists of 50 rounds of 100 iterations
with the penalization factor $\lambda=2$ and learning rate $\eta=0.005$.
Before each round, the current value of the trajectory $\mathbf{x}$
is updated by the solution of the forward ODE problem by the explicit second-order Runge-Kutta method.

In the RL setup, the agent chooses the forces $(F_1, F_3)$ in $[-1, 1]$ every $\Delta t$ units of time from the state $(L_1, L_2)$.
We use the reward
\begin{equation} \label{e_3beads_reward}
  r_t = x_{c}(t) - x_{c}(t-\Delta t),
\end{equation}
thus the cumulative reward evaluates to $x_c(T)$, which is the quantity to be maximized.
Each RL episode ends when the simulation time reaches $T$.
We use $\tau=0.1$ and train the RL agent for 5\,000 episodes.

\begin{figure}
  \centering
  \includegraphics{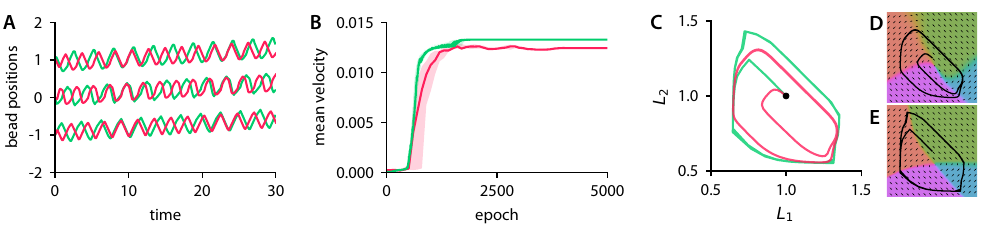}

  \caption{%
    Three beads swimmer trained using ODIL~\key{110} and RL~\key{120}.
    (\figbf{a})~Trajectory of the three beads.
    (\figbf{b})~Training history showing the mean velocity of the center of mass.
    (\figbf{c})~Lengths of the arms in the phase space starting at~\key{001}.
    (\figbf{d},~\figbf{e})~ODIL and RL policies in the phase space with arrows
    showing the vector $(-F_1,F_3)$ and colors showing its angle.
  }
  \label{f_3beads}
\end{figure}

The trajectories obtained with ODIL and RL are shown in~\cref{f_3beads}.
Both methods find similar strategies to propel the swimmer:
the arm lengths form a cycle in the $(L_1, L_2)$ space corresponding to non-reciprocal motion.
RL achieves a swimming velocity of $0.0133$, while ODIL finds a smaller value of $0.0124$.
We observe that the optimization of the ODIL loss function is sensitive to the
choice of hyperparameters: e.g. decreasing the factor $\lambda$ or using more iterations per round may reduce the obtained
swimming velocity.

\subsection{Path planning around an obstacle}
\label{s_obstacle}

Here we consider the problem of path planning around an obstacle in a high-dimensional space~\cite{huang2021cem}.
This case shows how ODIL and RL behave as the dimension of the state and action spaces increases.
A spherical obstacle of radius $R=0.25$ is centered at $\mathbf{x}_c = (0.5, 0, \dots, 0) \in \mathbb{R}^D$ in a $D$-dimensional space.
A particle $\mathbf{x}$ propels at a constant velocity $U=1$ without inertia and under a repulsion force that prevents the particle from penetrating the obstacle
\begin{equation} \label{e_obstacle_ode}
  \dot{\mathbf{x}} = U \mathbf{p} + 100 \max{(0, R-|\mathbf{r}|)} \frac{\mathbf{r}}{|\mathbf{r}|},
\end{equation}
where the unit vector $\mathbf{p}$ is the instantaneous direction and $\mathbf{r} = \mathbf{x} - \mathbf{x}_c$.
The task is to find $\mathbf{p}$ that brings the swimmer
from a starting position $\mathbf{x}_\text{start}=\mathbf{0}$
to a target position $\mathbf{x}_\text{target}=(1, 0, \dots, 0)$ in the minimal travel time~$T>0$.

In the ODIL setup, we use a grid size $N=129$ along the time coordinate.
The policy is a fully connected NN of size $128\times 128$
with input $\mathbf{x}$ and output $\mathbf{p}$.
The output is normalized so that $|\mathbf{p}|=1$.
The starting positions and targets are imposed exactly.
The optimization consists of 10 rounds of 1000 iterations,
with the penalization factor $\lambda=0.01$ and learning rate $\eta=0.001$ in the first five rounds that are multiplied by 0.25 before each of the subsequent five rounds.

In the RL setup we use the time limit $T_\text{max} = 5$, time penalty coefficient $\kappa=10$, time between actions $\tau = 0.02$, distance tolerance $\delta=0.05$, and train for 10\,000 episodes.
The integration time step in RL was set to $\Delta t = 0.02$.

\begin{figure}
  \centering
  \includegraphics{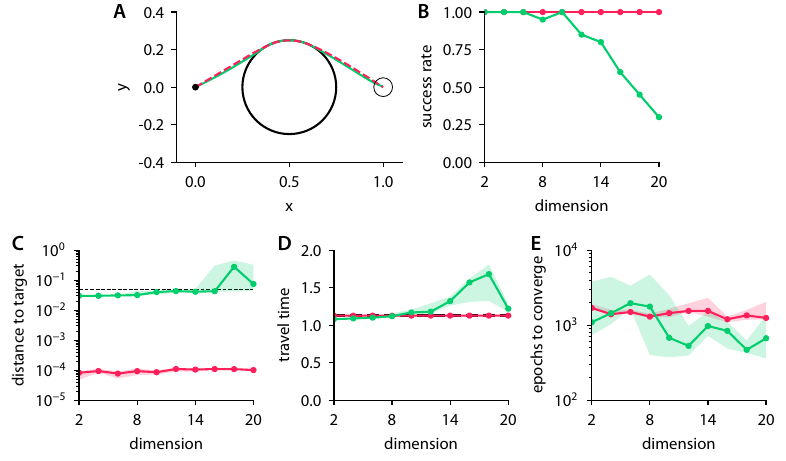}

  \caption{%
    Path planning around an obstacle.
    (\figbf{a})~Trajectories obtained for $D=2$ with ODIL~\key{210} and RL~\key{120}.
    The circle around the target has a radius $\delta$.
    (\figbf{b},~\figbf{c},~\figbf{d},~\figbf{e})~Success rate, distance to target, travel time, and number of epochs until convergence versus dimension $D$ from ODIL~\key{110} and RL~\key{120}.
    Shades are the $20^\text{th}$ to $80^\text{th}$ percentiles, solid lines are the medians over 10 realizations for RL and 20 realizations for ODIL.
  }
  \label{f_obstacle}
\end{figure}

To test the robustness of the two methods against the dimension of the states and actions, we vary $D=2,\;4,\dots,20$.
The results include 10 realizations with different random seeds.
\Cref{f_obstacle} shows the trajectories from both methods compared to the exact solution, achieved distance to target, success rate, number of epochs until
convergence, and travel time.
ODIL reliably finds a trajectory around the obstacle for all values of $D$, unlike RL that fails more than 50\% realizations for $D \geq 16$.
We explain this advantage of ODIL over RL by the information gained from the gradients of the loss function.
ODIL has access to the gradients through the governing equations,
while RL only observes a scalar rewards and relies on random sampling to update the policy.
The travel time of ODIL remains close to the exact solution, while the travel time of RL becomes larger as $D$ increases.

\section{Computational Cost}

\begin{table}
  \centering
  \begin{tabular}{l|c|c|c|c}
    \textbf{Case}             & \multicolumn{2}{c}{\textbf{Policy evaluations}} & \multicolumn{2}{c}{\textbf{Wall time (s)}} \\
                              & ODIL                     & RL              & ODIL                    & RL               \\ \hline
    brachistochrone           & \textbf{\num{1.29e+06}}  & \num{6.88e+07}  & \textbf{\num{1.12e+02}} & \num{6.95e+02}   \\
    path planning, shear      & \textbf{\num{1.29e+05}}  & \num{3.08e+06}  & \textbf{\num{1.28e+01}} & \num{1.65e+01}   \\
    path planning, vortices   & \textbf{\num{1.29e+06}}  & \num{4.02e+08}  & \textbf{\num{1.04e+02}} & \num{4.25e+03}   \\
    three-beads swimmer       & \textbf{\num{5.12e+06}}  & \num{1.16e+08}  & \textbf{\num{1.85e+02}} & \num{6.87e+02}   \\
    obstacle, $D=2$           & \textbf{\num{1.29e+06}}  & \num{3.00e+07}  & \textbf{\num{9.37e+01}} & \num{5.16e+02}   \\
    obstacle, $D=6$           & \textbf{\num{1.29e+06}}  & \num{8.31e+07}  & \textbf{\num{9.67e+01}} & \num{1.70e+03}   \\
  \end{tabular}
  \caption{
    Number of policy evaluations and total wall time for training the policy with ODIL and RL.
    The numbers correspond to the median over 10 random seeds (20 random seeds in the obstacle cases).
  } \label{tab:s_cost}
\end{table}

The cost of each example described in \cref{s_brach,s_obstacle,s_pathplan,s_threebeads} is evaluated with the same metrics as in the manuscript.
The results are summarized in \cref{tab:s_cost}.
In the case of the brachistochrone, the training is considered converged when the error \cref{e_brach_error} is less than 0.1.

\end{document}